# Nucleation kinetics in phase transformations with spatially correlated nuclei

Massimo Tomellini

Dipartimento di Scienze e Tecnologie Chimiche, Università di Roma Tor Vergata,

Via della Ricerca Scientifica 1, 00133 Roma, Italy

**Abstract**

Phase transitions ruled by nucleation and growth can occur by nonrandom arrangement of nuclei. This is verified, for instance, in thin film growth at solid surfaces by vapor condensation or by electrodeposition where, around each nucleus, a depletion zone of reactants sets up within which nucleation is prevented. In this contribution, a theoretical approach for the kinetics of phase transition with spatially correlated nuclei by progressive nucleation is developed. The work focuses on the rate of formation of the actual nuclei, a quantity that is necessary for describing the transformation kinetics. The approach is based on correlation functions and applied to treat hard-sphere interaction between nuclei. Computations have been performed for 2D and 3D growths by truncation of the series expansion in correlation functions up to second order terms. It is shown that the nucleation kinetics undergoes a transition from a typical Random Sequential Adsorption (RSA) behavior to one that is like the Kolmogorov–Johnson–Mehl–Avrami (KJMA) kinetics. The time evolution of the volume fraction of the new phase is found to depend slightly on correlation radius. Such behavior is explained by the partial balancing between the reduction in nucleation density and the decrease in impingement events, which have opposite effects on the kinetics.

**1-Introduction**

The model developed by Kolmogorov Johnson Mehl and Avrami (KJMA) [1-3] is an important theoretical tool in Materials Science for describing the kinetics of phase transformations by nucleation and growth [4-9]. Although the model was originally formulated in the ambit of phase transitions at the solid state, since then it has been successfully applied to a variety of fields, which include, just to cite a few, Biology, Cosmology, Social Science and Economics [10 and references therein].

It is well known that the theory is just an application of Poisson statistics to the nucleation process, since the spatial distribution of nucleation centres is assumed random throughout the whole space. In this case, the exact solution of the phase transformation kinetics is given by simple analytical expressions, useful for interpreting experimental data [4,11,12]. However, in several systems the Poissonian distribution of nuclei is the exception rather than the rule. A paradigmatic example is the electrodeposition of a new phase on solid substrate driven by progressive nucleation, where the spatial distribution of nuclei is not random [13-16]. This is due to the diffusion fields around growing nuclei implying a reduction of supersaturation and, consequently, of the nucleation probability around the already formed nuclei [17,18,19]. It stems that around each nucleus there is a depletion region where nucleation is forbidden; the arrangement of nuclei centres resembles a system of dots with hard-disk interaction [13]. Another process, important for technological application, is the growth of diamond film by seeding of the substrate surface with diamond nuclei [20,21]. Due to the finite size of the seeds, the formation of the diamond phase can be described as a phase transformation with site-saturated nucleation and hard-disk correlation between nuclei. Theoretical modeling with application to experimental data are reported in ref.[21].

During the last years, to extend the range of applicability of the theory, approaches have been formulated for describing phase transformation kinetics with correlated nuclei [22-25]. These modelling make use of the *n*-dots correlation functions for computing the probability that no dot is located within a given region of space. The theory has been applied to phase transitions with site-saturated nucleation [22,23,26] and generalized to the case of progressive nucleation in refs.[24,27]. At odds with the KJMA model, in the case of progressive nucleation of spatially correlated nuclei, the mathematical solution of the kinetics is expressed in terms of actual nucleation rate, rather than that comprehensive of phantoms [28]. The model case of transformations with correlated nuclei and constant value of the actual nucleation rate has been discussed in ref.[27]. A more realistic approach would require computing the actual nucleation rate as a function of the degree of advancement of the transformation, namely the time. The relationship between actual nucleation rate and time, $I_a(t)$, is quite evident for the classical KJMA kinetics for which,

$$I_a(t) = I_0\big(1 - \xi(t)\big), \quad (1)$$

with $\xi(t)$ being the volume fraction of the transformed phase at time $t$ and $I_0$ (taken here as constant) the nucleation rate throughout the whole space, with the inclusion of phantoms. It is worth pointing out that the rate $I_a$ is the number of actual nuclei formed in unit time divided by the total volume of the system. Consequently, $I_0$ is also equal to the rate of formation of actual nuclei per unit of *untransformed* volume. Eqn.1 holds true also for the parabolic growth law, although such a growth mode is non-compliant with the KJMA approach owing to the overgrowth of phantoms [29]. In this case the volume fraction is not given by the usual KJMA kinetics [30,31]. In fact, the computation of $\xi(t)$ for random nucleation and parabolic growth requires employing, in the stochastic approach, the actual nucleation rate given by eqn.1 and the correlation functions [32,33].

Because of the correlation constraint, in transformations with correlated nuclei eqn.1 does not hold, in general, since nucleation could be inhibited even in the untransformed volume. This point is made clear through Fig.1 which illustrates the case of progressive nucleation in 2D for both random distribution of nuclei (panels *a*, *b*) and for nuclei with hard-core correlation (panel *c*). As displayed in panel *c*), the hard-core correlation prevents nucleation even in the untransformed portions of the surface.

To get a deeper insight into this topic, in the present work we present an approach for computing the rate of appearance of *actual* nuclei in phase transformations with spatially correlated nuclei, according to the hard-sphere interaction. To the best of author's knowledge, the present modelling is new since previous contributions were devoted to the case of site-saturated nucleation or to eliminate phantom overgrowth [22,23,26,31,33]. Besides, the hard-core interaction is a benchmark in phase transition with correlated nuclei. It is a suitable model for describing nucleation in thin film growth by seeding and in electrochemical phase formation [13,21], as reported above.

The paper is divided as follows: in section 2.1 we outline the stochastic approach for correlated nucleation that is employed, in section 2.2, for modelling nucleation rate in progressive nucleation in two- and three-dimensional space. The outputs of the numerical computation are discussed in section 2.3.

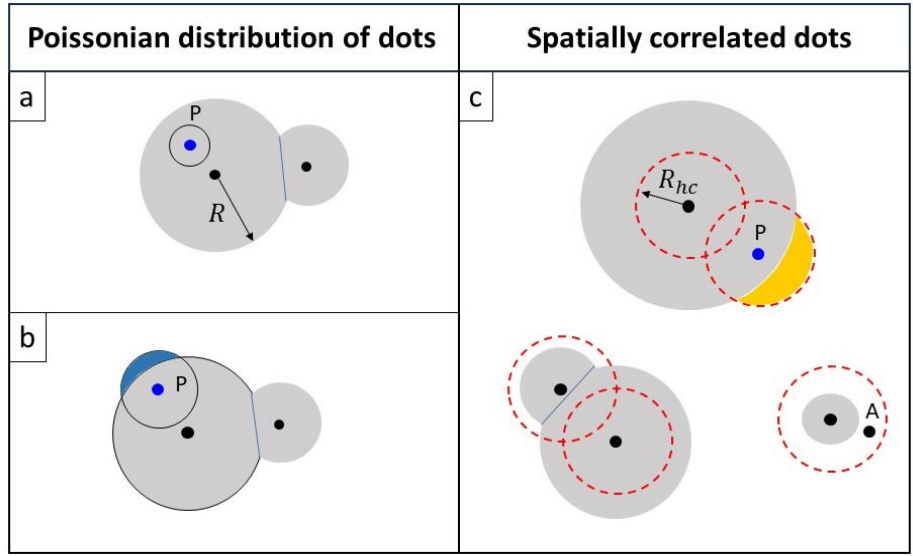

Fig.1 Pictorial view of an arrangement of dots (nuclei centres, black points) in phase transformation ruled by progressive nucleation and growth with impingement. In the figure, the blue dot, $P$, is a phantom, $R$ is the nucleus radius and $R_{hc}$ the correlation disk. Panel *a*) shows two actual nuclei and a phantom nucleus within the transformed phase. In the case of KJMA compliant growth (e.g. linear growth and random nucleation) the overgrowth phenomenon is precluded and the KJMA kinetics applies. Phantom overgrowth occurs for parabolic growth law, as depicted in panel *b*) (blue area of the phantom nucleus), where the KJMA kinetics does not hold for the $\xi(t)$. However, in both cases of panels *a*) and *b*) the actual nucleation rate is given by eqn.1.

Panel *c*): non-random nucleation according to the hard-disk correlation. In the figure, the dashed red circle is the correlation disk, within which nucleation is forbidden. At odds with the KJMA model, considering phantoms will alter the stochastic process, since nucleation could be precluded even where it should be allowed. This is the region colored in yellow within the correlation disk of the phantom. Nucleation can be also precluded in the uncovered portion of the surface, namely within the correlation disk of an actual nucleus (for instance in point A). In this case eqn.1 does not hold.

## 2-Results and discussion

*2.1 Phase transformations with spatially correlated nuclei*

The stochastic approaches for modeling the kinetics of phase transformations by nucleation and growth relay on the determination of the probability that no dots (i.e. the centers of nuclei) lay within a given volume centered at a generic point of space. In kinetics with site saturated nucleation and spherical nuclei the measure of this region, $|\Delta_t|$, is the volume of the nucleus at time $t$: $|\Delta_t| = \int_{\Delta_t} d\boldsymbol{r}$. From this probability the volume fraction of the transformed phase is easily obtained according to, $\xi(t) = 1 - Q_0(\Delta_t)$ where $Q_0(\Delta_t)$ is the probability that no dots are within the sphere $\Delta_t$ centered at a generic point of the space. In fact, the absence of any nucleus in a sphere of radius equal to the nucleus radius, at $t$, guarantees that the center of this sphere is not within any nucleus at that time, i.e. the generic point of the space does not belong to the new phase at $t$. For a random distribution of $N$ dots, the Poisson distribution gives $Q_0(\Delta_t) = e^{-N|\Delta_t|}$, that leads to the KJMA kinetics [1]. In the case of spatially correlated nuclei with site-saturated nucleation, the above probability is given in terms of a series of correlation functions. Truncation of the series up to the second order term provides [22,23]

$$Q_0(\Delta_t) = \exp\left[-N|\Delta_t| + \frac{1}{2}N^2 \int_{\Delta_t} d\boldsymbol{r}_1 \int_{\Delta_t} d\boldsymbol{r}_2\, g_2(\boldsymbol{r}_1,\boldsymbol{r}_2)\right], \qquad (2)$$

where $g_2$ is the 2-dots correlation function that is related to the pair distribution function, $g(\boldsymbol{r}_1,\boldsymbol{r}_2)$, through the expression $g_2(\boldsymbol{r}_1,\boldsymbol{r}_2) = g(\boldsymbol{r}_1,\boldsymbol{r}_2) - 1$. For homogeneous systems, translationally invariant as considered here, $g(\boldsymbol{r}_1,\boldsymbol{r}_2) = g(|\boldsymbol{r}_1 - \boldsymbol{r}_2|) = g(r)$.

Eqn.2 has been generalized to the case of progressive nucleation implying, at any time, a distribution of nucleus size. To this purpose it is considered a set of distinguishable classes of nuclei (dots). Within each class dots are indistinguishable. Since nuclei start growing at different times, the growth law now depends on both actual time, $t$, and birth time, $t'$, of the nuclei. The classes of dots are labelled with the nucleus birth times to which, in this case, the correlation function may also depend. For instance, the correlation function of a couple of nuclei with birth times $t_1$ and $t_2$ reads $g_2(\boldsymbol{r}_1, \boldsymbol{r}_2, t_1, t_2)$. Employing the same order of approximation of eqn.2, the probability that no dots of the "$t'$-class" is located within the domain $\Delta_{t',t}$, for any $t'$ in the range $0 < t' < t$, becomes [27] (see also Appendix A)

$$Q_0(\Delta_{0,t}) = \exp\left[-\int_0^t I_a(t')|\Delta_{t',t}|\,dt' + \int_0^t I_a(t_1)\,dt_1 \int_0^{t_1} I_a(t_2)dt_2 \int_{\Delta_{t_1,t}} d\boldsymbol{r}_1 \int_{\Delta_{t_2,t}} d\boldsymbol{r}_2\, g_2(\boldsymbol{r}_1,\boldsymbol{r}_2,t_1,t_2)\right], \qquad (3)$$

where $I_a(t')$ is the nucleation rate of actual nuclei at time $t'$ and $\Delta_{0,t}$ stands for the whole sequence of the $\Delta_{t',t}$ domains. The volume fraction of the transformed phase is given by eqn.3 by considering for $\Delta_{t',t}$ the sphere with radius equal to the nucleus radius at $t$. In fact, by denoting with $R(t,t') \equiv R(t-t')$ the radius of the nucleus, we get $|\Delta_{t',t}| = \int_{\Delta_{t',t}} d\boldsymbol{r} = \nu_D R(t,t')^D$ with $D$ space dimension and $\nu_D$ a geometrical factor ($\nu_2 = \pi$, $\nu_3 = 4\pi/3$). In other words, $Q_0(\Delta_{0,t})$ is the probability that no nucleus born in time interval $0 < t' < t$ is within a sphere of radius $R(t,t')$ centered at a generic point of the space. It follows that $Q_0(\Delta_{0,t})$ is the probability that a generic point of space does not belong to the new phase, from which the volume fraction is obtained: $\xi(t) = 1 - Q_0(\Delta_{0,t})$.

When $I_a(t')$ is given by eqn.1, eqn.3 provides an integral equation for $\xi$ or, equivalently, for $\frac{I_a(t')}{I_0} = Q_0(\Delta_{0,t'})$. In this case, the distribution of dots is random through the space, yet correlation is present between the *actual nuclei*. In fact, the relative distance between two nuclei, born at $t_1$ and $t_2$, must satisfy the constraint $r > R(t_1 - t_2)$ for nucleus 1 not to be a phantom [33]. Therefore, the radial distribution function is nil for $r < R(t_1 - t_2)$ which implies spatial correlation among actual nuclei even in KJMA compliant growths. This point is further discussed in section 2.2. Eqn.3 was employed in refs.[28,33] to avoid phantom overgrowth in parabolic growths.

*2.2 Nucleation rate of actual nuclei*

As anticipated in the introduction, eqn.2 holds whenever nuclei form at random in the untransformed portion of the volume. However, as illustrated in Fig.1c, spatial correlation among nuclei may prevent nucleation even in the untransformed phase, for which eqn.1 is invalid. The computation of the rate of formation of actual nuclei is more involved, in this case, since requires tackling a different stochastic process than the one for the $\xi(t)$ function (eqn.3).

In the following, this topic is studied for the hard-core interaction between actual nuclei where, at the lowest order, the pair distribution function of dots is equal to $g(r) \cong H(r - R_{hc})$, with $H(\cdot)$ the Heaviside step function and $R_{hc}/2$ the core radius. In fact, higher order terms in the series expansion of the $g(r)$ can be neglected for low densities of nuclei. The validity of this approximation in phase transformation has been discussed in ref.[22]. However, the $g(r)$ above reported only holds for nucleation that could occur within the correlation sphere of another nucleus. In fact, given two nuclei with birth times $t_1$ and $t_2$ (where $t_2 < t_1$ is considered) the conditions according to which both their distance is greater than $R_{hc}$ and the first nucleus is not a phantom, requires the constraints $r > R_{hc}$ and $r > R(t_1 - t_2)$ to be satisfied. Consequently, the pair distribution function is given by:

$$g(r, t_1, t_2) = \begin{cases} H(r - R_{hc}), & \text{for } R(t_1 - t_2) < R_{hc} \\ H(r - R(t_1 - t_2)), & \text{for } R(t_1 - t_2) > R_{hc} \end{cases}, \quad (4)$$

where $|\mathbf{r}_1 - \mathbf{r}_2| = r$ is the relative distance between the couple of nuclei with birth time $t_1$ and $t_2$.

The stochastic process for computing $I_a$ can be envisaged according to the following: between time $t$ and $t + \Delta t$, $I_0 \Delta t$ dots, distributed at random in the whole volume, start growing with an exception for the dots, located either within the new phase or closer to another nucleus than $R_{hc}$, that are removed from the system. The actual nucleation rate is therefore given by

$$I_a(t) = I_0 Q_a(t), \quad (5)$$

where $Q_a$ is the probability that a generic point of the system does not belong to the new phase and its distance from next neighbour nuclei exceed $R_{hc}$. This probability is computed by means of eqn.3 with a judicious choice of the $\Delta_{t',t}$ integration domain. We recall that eqn.3 provides the probability that no dot lays in this domain: in fact, in the $g(r)$ containing term, the $\Delta_{t_1,t}$ and $\Delta_{t_2,t}$ refer to the exclusion of dots 1 and 2, respectively, from these domains (see also Appendix A). In the following, we consider linear growth of nuclei, $R(t, t') = a(t - t')$ with $a$ the growth rate. To determine the $\Delta_{t',t}$ domain we define the time $t^*$ from the equation $R(t, t^*) = R_{hc}$, namely $t^*(t) = t - \frac{R_{hc}}{a}$ at running time $t$. For $t > \frac{R_{hc}}{a}$ the time $t^*$ defines two populations of nuclei with sizes greater and shorter than the correlation radius, $R_{hc}$. This is shown in Fig.2 where the growth law is plotted as a function of running time and for several values of the birth time, $t'$, in the interval $(0, t)$. Based on this, for the present stochastic process the radius of the spherical domain $\Delta_{t',t}$ is

$$r(t, t') = [R_{hc} H(t' - t^*) + R(t, t') H(t^* - t')] H\left(t - \frac{R_{hc}}{a}\right) + R_{hc} H\left(\frac{R_{hc}}{a} - t\right), \quad (6)$$

that will be employed in eqn.3 for transformations in both 2D and 3D space.

The computation of eqn.3 with the pair distribution function eqn.4 requires particular attention, depending on both the birth times of the couple of nuclei, $t_1$ and $t_2$ (where $t_2 < t_1$ and $R(t, t_2) > R(t, t_1)$) and on the size of the second nucleus at the birth time of the first, $R(t_1, t_2) = R(t_1 - t_2)$.

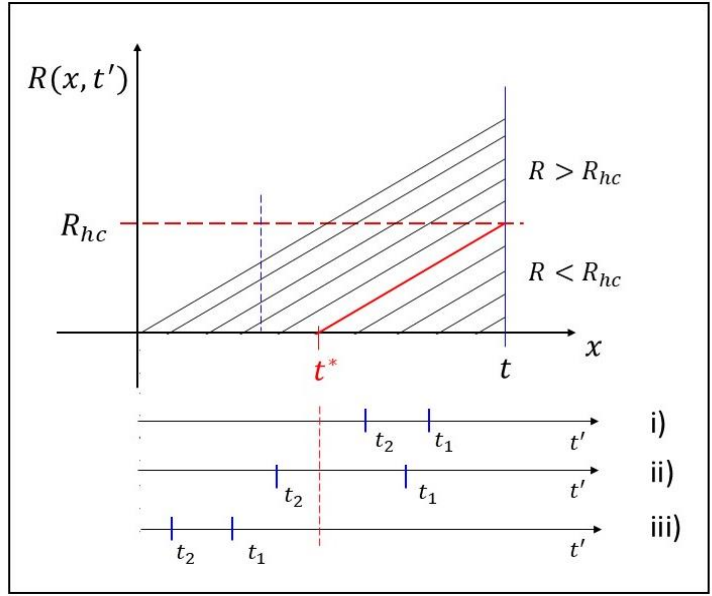

Fig.2 Linear growth law of nuclei for several values of the birth time, $t'$, in the range $0 < t' < t$. The growth law is $R(x, t') = a(x - t')$ with $x \geq t'$. At running time $t > \frac{R_{hc}}{a}$, where $R_{hc}$ is the hard sphere radius, two sets of nuclei are defined, with $R > R_{hc}$ and $R < R_{hc}$, namely with birth time shorter or larger that $t^*(t) = t - \frac{R_{hc}}{a}$. The figure also displays the birth times of the couple of nuclei considered in the second order contribution of the probability function, i.e. for the cases i)-iii) of eqns.7a-c. At time $t < \frac{R_{hc}}{a}$ (marked by a dashed blue line in the graph), there is just one set of nuclei with $R < R_{hc}$ and the stochastic process is the same as the RSA.

According to eqns.4,6 the following cases are considered ($t^* > 0$ i.e. $t > \frac{R_{hc}}{a}$):

a- for $R(t_1 - t_2) < R_{hc}$:

i)     $t^* < t_2 < t_1$,     $r(t, t_1) = r(t, t_2) = R_{hc}$;                                                            (7a)

ii)     $t_2 < t^* < t_1$,     $r(t, t_1) = R_{hc}$, $r(t, t_2) = R(t, t_2)$;                              (7b)

iii)    $t_2 < t_1 < t^*$,     $r(t, t_1) = R(t, t_1)$, $r(t, t_2) = R(t, t_2)$.                        (7c)

b- for $R(t_1 - t_2) > R_{hc}$:

the cases ii) and iii) above still hold whereas case i) does not. In fact, case i) implies

$t_1 - t_2 < t - t^*$ and $a(t_1 - t_2) = R(t_1 - t_2) < a(t - t^*) = R_{hc}$, that is incompatible with the constraint $R(t_1 - t_2) > R_{hc}$.

For $t < \frac{R_{hc}}{a}$ all growing nuclei are within the correlation sphere and the stochastic problem of finding $Q_a(t)$ is equivalent to that of the Random Sequential Adsorption (RSA), for the random packing of spheres of radius $R_{hc}/2$ in D-dimensional space [34-36]. This point is further discussed in the next section.

As regards the spatial integral over the correlation function entering eqn.3, it can be simplified using polar or spherical coordinates. Since the system is homogeneous, we get:

$$\psi(t, t_1, t_2) = \int_{\Delta_{t_1,t}} d\boldsymbol{r}_1 \int_{\Delta_{t_2,t}} d\boldsymbol{r}_2\, g_2(\boldsymbol{r}_1, \boldsymbol{r}_2, t_1, t_2)$$

$$= \Omega_D \int_0^{R(t_1,t)} r_1^D dr_1 \int_{\Delta_{t_2,t}} d\boldsymbol{r}_2\, g_2(r, t_1, t_2), \qquad (8)$$

where $\Omega_D$ is the solid angle ($\Omega_3 = 4\pi$) or the polar angle ($\Omega_2 = 2\pi$) and $r = |\boldsymbol{r}_1 - \boldsymbol{r}_2|$ the relative distance. Since $g_2(r, t_1, t_2) = g(r, t_1, t_2) - 1$, the integral provides two second order terms, i.e. due to the Heaviside function and to the $-1$ constant (eqn.4). The integral with the Heaviside function can be expressed in terms of the overlap volume of two spheres of radius $R(t, t_2)$ and $R_{hc}$, for $R(t_1 - t_2) < R_{hc}$, and of radius $R(t, t_2)$ and $R(t_1 - t_2)$ for $R(t_1 - t_2) > R_{hc}$ [33]. In eqn.8 the second order contribution due to the term $-1$ is equal to $-|\Delta_{t_1,t}||\Delta_{t_2,t}|$. Finally, the $\psi(t, t_1, t_2)$ function (eqn.8) is integrated with the actual nucleation rate over the time variables $t_1$ and $t_2$. Fig.3 shows the integration domains of the double integral in time, for the cases i)-iii) of eqn.7a-c. In the figure, the different colors highlight that for these cases the spatial integral, $\psi(t, t_1, t_2)$, are in general different. The straight line, in red, is the equation $t_2 = t_1 - \frac{R_{hc}}{a}$ (i.e. $R(t_1 - t_2) = R_{hc}$) at the boundary between conditions $R(t_1 - t_2) < R_{hc}$ and $R(t_1 - t_2) > R_{hc}$ whose domains are above and below this line, respectively. For $t < \frac{R_{hc}}{a}$ the red line shifts to the dashed blue one and only the condition $R(t_1 - t_2) < R_{hc}$ holds true. This is equivalent to the RSA. Moreover, for $R_{hc} = 0$, the KJMA model is recovered and the integration domain is that for case

iii) with $R(t_1 - t_2) > R_{hc}$ (white colored domain in Fig.3). However, because of the correlation between actual nuclei eqn.8 is different from zero even for $R_{hc} = 0$.

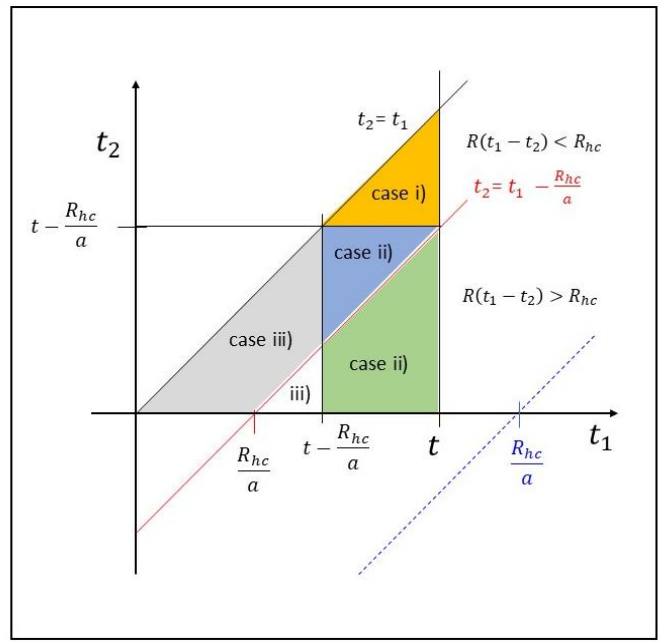

Fig.3 Integration domains of the time integral according to the constraints reported in eqns.7a-c. Since $t_1 > t_2$, the integration domain is below the line $t_2 = t_1$. Similarly, the conditions $R(t_1 - t_2) < R_{hc}$ and $R(t_1 - t_2) > R_{hc}$ are satisfied above and below the red line $t_2 = t_1 - \frac{R_{hc}}{a}$, respectively. The different colors of the domains imply different time-functions resulting from the space integral. The graph shows the case $t > \frac{R_{hc}}{a}$ when the maximum size is greater than the correlation sphere.

For the case $t < \frac{R_{hc}}{a}$ the equation $t_2 = t_1 - \frac{R_{hc}}{a}$ is displayed as dashed blue line. The condition $R(t_1 - t_2) > R_{hc}$ is not satisfied and the stochastic process for determining $Q_a$ is the same as the RSA.

## 2.3 Numerical computation of the actual nucleation rate and volume fraction

This section is devoted to the numerical computation of both nucleation kinetics and volume fraction of the new phase for 2D and 3D transformations. The numerical computations have been carried out using the Wolfram Mathematica package.

The rate of formation of actual nuclei is computed through eqns.3,4 with the integration domains reported in eqns.7a-c and in Fig.3. Use of eqn.3 in eqn.5 results in an integral equation for $I_a(t)$. From the knowledge of $I_a(t)$ it is possible to determine the volume fraction of the transformed phase through eqn.3, in which $|\Delta_{t',t}| = \frac{\Omega_D}{D} R(t, t')^D$ is the nucleus volume [27].

In the computation, we used reduced quantities where lengths are normalized to the maximum size of the nucleus, $R(t,0) = at$. In the case of linear growth, as considered here, we define $\rho_{hc} = \frac{R_{hc}}{at}$ and the reduced time $\tau^* = \frac{t^*}{t} = 1 - \rho_{hc}$ for $\rho_{hc} < 1$. With the same normalization the growth law becomes $\rho(\tau_i) = \frac{a(t-t_i)}{at} = 1 - \tau_i$ (with $i = 1, 2$) and the $r_1$ variable changes in $x_1 = \frac{r_1}{at}$. The rate $I_0$ is constant and, in analogy with the KJMA kinetics, the extended volume is given by [1] $X_{ex}(t) = \frac{\Omega_D}{D(D+1)} I_0 a^D t^{D+1} = \left(\frac{t}{\tau_D}\right)^{D+1}$, namely $V_{ex} = \left(\frac{t}{\tau_3}\right)^4$ and $S_{ex} = \left(\frac{t}{\tau_2}\right)^3$, respectively for transformations in 3D and 2D. In terms of reduced time, $\bar{t} = \frac{t}{\tau_D}$, one gets $\rho_{hc} = \frac{1}{\bar{t}}\left(\frac{\gamma}{3}\right)^{1/(D+1)}$ where

$$\gamma = 3 R_{hc}^{D+1} \frac{\Omega_D}{D(D+1)} \frac{I_0}{a} = R_{hc}^{D+1} \frac{\pi I_0}{a}$$ is a measure of the degree of correlation between nuclei. For the KJMA model $\gamma = 0$.

Eqns.3,4 provide the following integral equation for $Q_a(\tau)$,

$$Q_a(\bar{t}) = \exp(F[Q_a]), \qquad (9)$$

where $F[Q_a]$ denotes the integral expression containing the $Q_a$ probability function reported in eqn.B1 in Appendix B.

To test the validity of the present approach based on eqns.3,4, we first apply eqn.9 to KJMA compliant transformations for which $Q_a(\bar{t})$ is given by eqn.1 according to $Q_a(\bar{t}) = e^{-X_{ex}(\bar{t})} = e^{-(\bar{t})^{D+1}}$. Specifically, $F[Q_a]$ (eqn.B1 in Appendix) has been computed, as a function of $\bar{t}$, at $\rho_{hc} = 0$ by using $\tau_i = \frac{t_i}{t}$ ($i = 1,2$) as integration variables and $Q_a(\tau_i) = \exp\left[-\left(\frac{t\tau_i}{\tau_D}\right)^{D+1}\right] = \exp[-(\tau_i)^{D+1} X_{ex}(\bar{t})]$ [2]. Next, the output of the computation has been compared with the solution of the KJMA theory to check the validity of eqn.9, namely the fulfillment of the identity $e^{-(\bar{t})^{D+1}} = \exp(F[\exp(-(\tau)^{D+1} X_{ex}(\bar{t}))])$.

As displayed in Fig.4, there is a very good agreement between the exact solution and the actual nucleation rate, eqn.3, for 3D and 2D transformations. In the case of spatially correlated nuclei,

---

[1] For constant nucleation rate, $I_0$ (with the inclusion of phantoms) the extended volume is defined as:
$X_{ex} = I_0 \int_0^t v(t,t')dt'$, where $v(t,t')$ is the nucleus volume.
[2] To simplify the notation, we retain the same symbol for either $Q_a(\tau_i)$ and $Q_a(t_i)$ functions, as well as in similar occurrences in the text.

eqn.9 has been solved numerically by successive iterations and for several values of $\bar{t}$: $Q_a^{(k)} = \exp\left(F[Q_a^{(k-1)}]\right)$, with $k \geq 1$ number of iterations. For the function $Q_a^{(0)}$ we have chosen the KJMA solution: $Q_a^{(0)}(\bar{t}) = e^{-X_{ex}(\bar{t})}$. In general, with about 10 iterations the relative integral error is of the order of 0.02% at the highest correlation value investigated here. Typical behavior of the error with the number of iterations is reported in the Appendix B (Fig.B1).

The time dependence of the actual nucleation rate is displayed in Fig.5 for several values of $\gamma$.

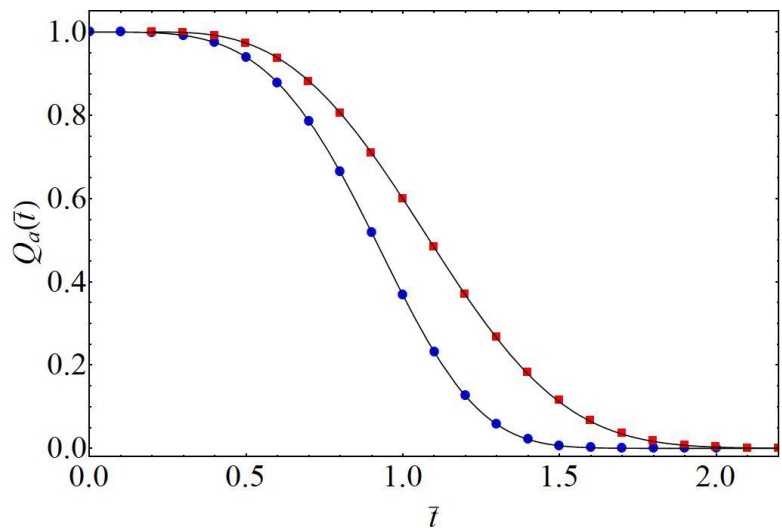

Fig.4 Exact analytical solution of the kinetics for random distribution of nuclei according to eqn.1 and the KJMA theory: $Q_a = \exp[-(\bar{t})^{D+1}]$ (solid lines for $D = 2$ and $D = 3$). The plot also shows the kinetics computed through eqn.9 at $\gamma = 0$, with $Q_a$ given by the KJMA solution above reported (symbols). Solid circles and solid squares refer to transformations in 3D and 2D, respectively. The kinetics at $D = 2$ has been shifted by $\Delta\bar{t}$=0.2. The integral error is about 0.04% and 0.3% for $3D$ and $2D$ transitions, respectively. The very good agreement between the two sets of data gives support to the second order approximation employed in the computation.

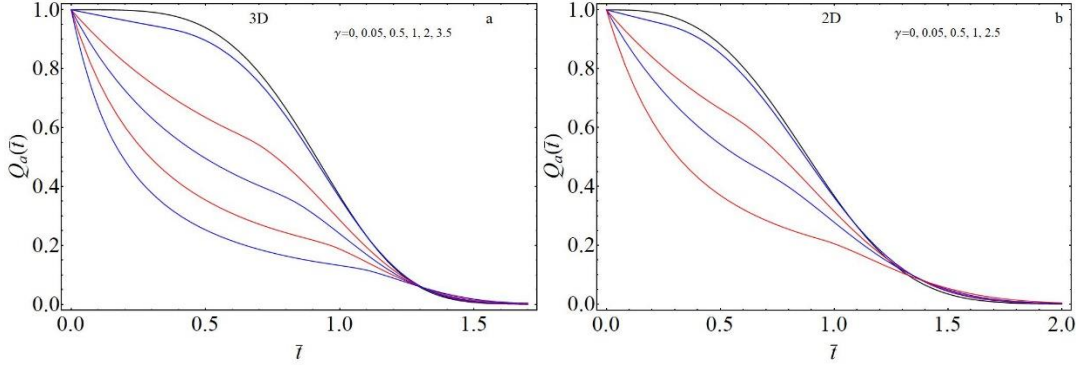

Fig.5 The nucleation rate, $Q_a = \frac{I_a}{I_0}$, is displayed as a function of reduced time, $\bar{t}$, for several values of $\gamma$ for 3D (panel *a*) and 2D (panel *b*) transformations. The correlation degree, $\gamma$, increases along the set of curves from top to bottom.

The hard-core correlation imparts a change in the curvature of the $\frac{I_a}{I_0}$ vs $\bar{t}$ function in the initial portion of the kinetics that is wider the higher $\gamma$. In fact, the initial behavior of the function resembles the RSA kinetics that is completely recovered up to $t_{hc} = \frac{R_{hc}}{a}$, i.e. $\rho_{hc} = 1$, from which $\bar{t}_{hc} = \left(\frac{\gamma}{3}\right)^{1/(D+1)}$. This value is close to the first inflection point of the curves of Fig.5a that occur at $\bar{t} = 0.3, 0.6, 0.7, 0.85$ and $0.95$ with increasing $\gamma$, in fair agreement with the values $\bar{t}_{hc} = 0.36$, $0.64, 0.76, 0.9$ and $1.0$. For the curves in Fig.5b the inflection points are at $\bar{t} = 0.2, 0.52, 0.67$ and $0.92$ in fair agreement with the values $\bar{t}_{hc} = 0.25, 0.55, 0.69$ and $0.92$. After these times, the curves get closer to the KJMA kinetics the lower $\gamma$ and eventually merge to the tail of the Poissonian curve.

Fig.6 displays the density of actual nuclei with time for the dimensionless quantity $N(\bar{t}) = \left(\frac{a}{I_0}\right)^{D/(D+1)} n_a(\bar{t})$, where $n_a(\bar{t}) = I_0 t(\bar{t}) \int_0^1 Q_a(\tau) d\tau$. The nucleation density increases with time, monotonically, to reach the asymptotic value that is higher the lower the hard-core radius (i.e. $\gamma$).

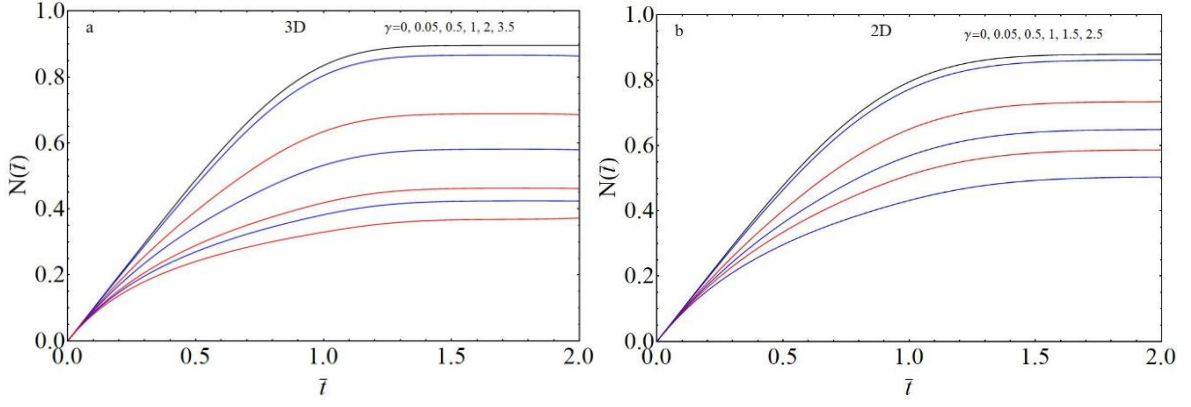

Fig.6 Nucleation density of actual nuclei as a function of reduced time, $\bar{t}$, and correlation degree ($\gamma$) for 3D and 2D transitions (panels $a$ and $b$). The actual nucleation density is normalized to the quantity $\left(\frac{I_0}{a}\right)^{D/(D+1)}$. In both panels the nucleation density at saturation decreases with increasing $\gamma$.

The nucleation density is further employed to compare the nucleation rate with that of the RSA process of spheres with radius $\frac{R_{hc}}{2}$. Let us consider the RSA process at time $t$, when $n_a(t)$ spheres have been placed into the system. At time $t$ the adsorption rate is given by $\frac{dn_a(t)}{dt} = I_0 Q_0(\Delta_{RSA}, n_a(t))$, where $\Delta_{RSA}$ is a sphere of radius $R_{hc}$ [34,35] and $Q_0(\Delta_{RSA}, n_a(t))$ the probability that no center of the $n_a(t)$ spheres is found within $\Delta_{RSA}$. Accordingly, to determine $Q_0(\Delta_{RSA}, n_a(t))$ we apply eqn.2 for the site-saturated nucleation with time dependent $N \equiv n_a(t)$.

The kinetics of RSA is studied as a function of the volume fraction that is occupied by the non-overlapping spheres, namely $\theta(t) = \frac{\Omega_D}{D}\left(\frac{R_{hc}}{2}\right)^D n_a(t)$. It is worth recalling that eqn.2 is suitable to study the adsorption kinetics, although it does not provide the exact value of the packing at the jamming point, $\theta_c$ [22,35]. In fact, $Q_0(\Delta_{RSA})$ must be zero for $\theta \geq \theta_c$, a condition that is not satisfied by the functional form of eqn.2 that approaches zero, asymptotically.

From eqn.2 the RSA kinetics is given by

$$Q_{RSA}(\theta) = \exp\left[-2^D \theta(1 + 2^{D-1}\theta) + \frac{D^2 2^{2D-1}}{\Omega_D}\theta^2 \beta_D\right], \qquad (10)$$

where $\beta_3 = \frac{17\pi}{72}$ and $\beta_2 = \frac{3\sqrt{3}}{8}$ are related to the mean value of the overlap area of two spheres of unitary radius. The comparison between the actual nucleation rate in phase transformations with correlated nuclei and the rate of RSA is displayed in Fig.7. In the figure, the probabilities $Q_{RSA}$ and $Q_a$ are plotted as a function of $\theta$, namely the total volume occupied by the adsorbed spheres of radius $\frac{R_{hc}}{2}$. In Fig.7 the dotted line is the RSA kinetics eqn.10 and the solid line the nucleation rate of the phase transformation. The kinetics do show that the nucleation rate coincides with the rate of the RSA up to the value $Q_a(\theta(\bar{t}_{hc}))$ marked on the y-axis. Beyond this point, the growth process mainly rules the kinetics and the shape of the curve becomes like the KJMA one, which is the nearly vertical line in Fig.7 (parallel to the $Q_a$ axes). In fact, for $R_{hc} = 0$, $\theta$ is identically nil. In panel $a$) it is also reported the behavior of the volume fraction with $\theta$, for two values of $\gamma$. It stems that the transformation reaches completion for

$\bar{t} \to \infty$, $Q_a \to 0$ and $\theta \to \theta(\infty) = n_a(\infty)\frac{\Omega_D}{D}\left(\frac{R_{hc}}{2}\right)^D$, where $n_a(\infty)$ is the nucleation density at saturation (Fig.6). Fig.7 also shows that, although the nucleation density at saturation decreases with $\gamma$, $\theta(\infty)$ increases with correlation radius and its value is always lower than the value at the jamming point. This is due to the growth process because adsorption of the spheres is not allowed even in some portions of the new phase.

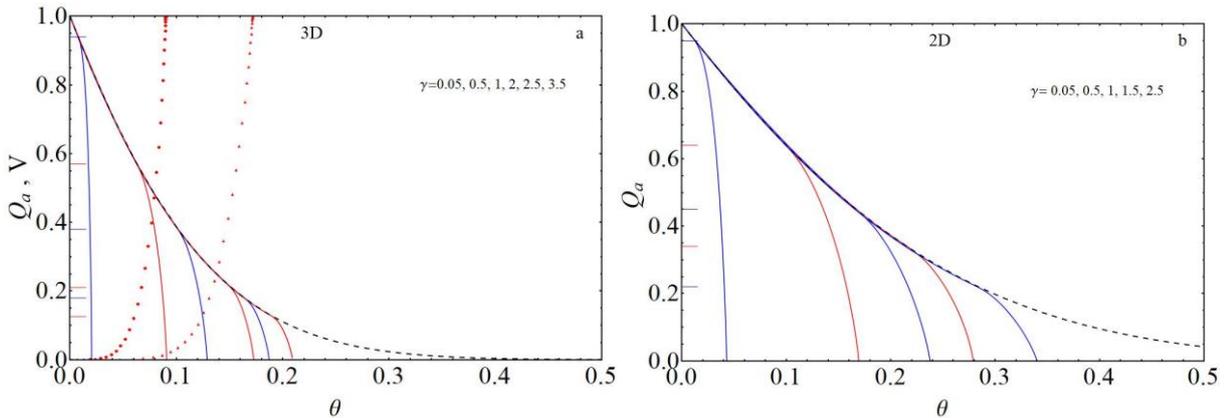

Fig.7 Comparison between the kinetics of nucleation in phase transformation with hard-core correlation and RSA, for 3D (panel $a$) and 2D (panel $b$) transformations. The dashed line is the RSA kinetics computed through eqn.10, where $\theta$ is the fraction of volume that is occupied by the spheres of radius $\frac{R_{hc}}{2}$. In the curves, the $\gamma$ value increases from left to right. The kinetics follow the RSA curve up to time $\bar{t}_{hc}$ when $\rho_{hc} = 1$. At this point, the growth of nuclei rules the kinetics that deviate from the RSA curve. The transition points are marked on the $Q_a$ axis for the various curves. After these points, the kinetics become like the KJMA

behavior that is the vertical line parallel to the $Q_a$ axis. Panel *a*) also shows the behavior of the volume fraction, $V(\theta)$, for $\gamma = 0.5$ and $\gamma = 2$ (symbols). The asymptotic values, $\theta(\infty)$, correspond to zero nucleation rate, $Q_a = 0$.

As explained in section 2.1, once evaluated, the actual nucleation rate is used in eqn.3 to determine the volume fraction of the transformed phase. The outcome of the computation is reported in Fig.8 for 3D (panel *a*) and 2D (panel *b*) transitions. In both panels an increase in the correlation strength, $\gamma$, causes a slight slow-down of the kinetics. In fact, the Avrami exponents of these kinetics are quite close to the values of the random case, $n = 4$ and $n = 3$ for 3D and 2D, respectively. Specifically, for the curves of panel *a*) it is in the range $3.92 < n < 4$ and for panel *b*) in the interval $2.9 < n < 3$. Such behavior can be understood by considering the opposite role played by both nucleation and impingement between nuclei on the kinetics. On one hand, the higher the correlation degree the lower the nucleation density and the rate of transformation (Fig.6). On the other hand, at higher correlation the overlaps between nuclei diminish, since nuclei are more isolated due to the greater volume within which nucleation is prevented. Consequently, an unimpeded growth of nuclei increases the rate of transformation. The results of Fig.8 indicate that the effect of the reduction in nucleation density is a little more important than that due to impingement. In fact, the small displacement of Avrami's exponent from the random case points to a nearly balancing of the two effects.

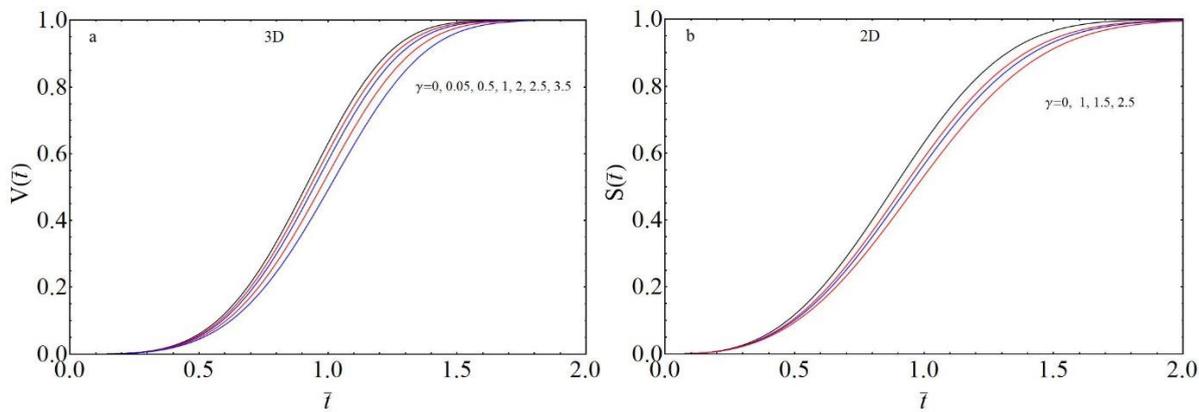

Fig.8 Kinetics of phase transformations with correlated nucleation in 3D (panel *a*) and 2D (panel *b*) space. The kinetics for linear growth of spherical nuclei have been computed through eqn.3 using the nucleation rate of Fig.3 (section 2.2). In the curves, the $\gamma$ value increases from top to bottom.

The weak dependence of $V(t)$ and $S(t)$ on $\gamma$ entails that these are not the most appropriate experimental quantities to highlight correlation effects between nuclei in phase transitions. This conclusion has been previously reached in ref.[25] based on computer simulations. On the other hand, according to Figs.5,6 the impact of correlation on nucleation rate (and density) is more important; this makes these quantities more suitable for experimental studies on correlation effect in phase transformations [13].

### 3-Conclusions

In this work, we proposed a method for modeling the nucleation rate of actual nuclei in phase transitions with spatially correlated nuclei. The case of hard-sphere interaction and linear growth of nuclei has been studied. It is shown that the nucleation rate is the solution of an integral equation solved by successive iterations. To check the validity of the second order approximation, on which the method rests, the approach has been applied to KIMA compliant transformations, for which the exact solution is known. A very good agreement is obtained between KJMA solution and the second order expansion of the kinetics in terms of correlation functions.

The kinetics of nucleation has been studied as a function of the correlation radius, $R_{hc}$, in 2D and 3D. It is shown that the kinetics exhibits two regimes that are more distinct the larger the correlation degree. In the first, the kinetics is ruled by the correlated nucleation, like in the RSA process. In the second, the kinetics is ruled by the growth of the new phase and resemble the kinetics of the KJMA model.

The computation of the actual nucleation rate is required for determining the time evolution of the volume fraction of the new phase. This quantity has been evaluated by solving a different stochastic approach than the one employed for $\xi(t)$, still using correlation functions at the same order of approximation. It is found that the rate of transformation decreases slightly with correlation radius. Such behavior has been ascribed to the reduction in nucleation density owing to correlation effects. In fact, the larger the correlation radius the larger the volume where nucleation is inhibited. Therefore, the fact that nuclei are more isolated in the nonrandom nucleation has a minor impact on the kinetics when compared to the reduction in nucleation rate.

**Appendix**

*A - Derivation of eqn.3.*

Eqn.3 is derived by considering a set of different classes of dots. Dots belonging to different classes are distinguishable whereas dots belonging to the same class are indistinguishable. For two classes of dots the generalization of eqn.2 reads:

$$Q_0(\Delta_1, \Delta_2) = Q_0(\Delta_1)Q_0(\Delta_2)\exp\left[N_1 N_2 \int_{\Delta_1} d\boldsymbol{r}_1 \int_{\Delta_2} d\boldsymbol{r}_2 \, g_2^{(1,2)}(\boldsymbol{r}_1, \boldsymbol{r}_2)\right] \quad (A1)$$

where $N_1$ and $N_2$ are the density of the two sets of dots and $Q_0(\Delta_i)$ ($i = 1,2$) is given by eqn.2. $Q_0(\Delta_1, \Delta_2)$ is the probability that no dots of class 1 and 2 lay in the domains $\Delta_1$ and $\Delta_2$, respectively. For $m$ classes of dots with density $N_i$ ($i = 1,2,\ldots,m$) we get

$$Q_0(\Delta_1, \Delta_2, \ldots, \Delta_m) = \exp\left[-\sum_{i=1}^{m} N_i|\Delta_i| + \frac{1}{2}\sum_{i,j} N_i N_j \int_{\Delta_i} d\boldsymbol{r}_1 \int_{\Delta_j} d\boldsymbol{r}_2 \, g_2^{(i,j)}(\boldsymbol{r}_1, \boldsymbol{r}_2)\right], \quad (A2)$$

where the 2-dots correlation function depends on the couple $(i,j)$ and the factor ½ avoids double counting. The continuum limit of eqn.A2 is performed by labelling the classes with a continuous variable, say $t'$. The number of dots of the "$t'$-class" then becomes, $N_i \to \frac{dN(t')}{dt'}dt' = I_a(t')dt'$ with the integration domain $\Delta_{t',t}$, where $t$ is the actual time and $0 < t' < t$. Eqn.A2 eventually becomes

$$Q_0(\Delta_{0,t}) = \exp\left[-\int_0^t I_a(t')|\Delta_{t',t}|\,dt' \right.$$
$$\left. + \int_0^t I_a(t')dt' \int_0^{t'} I_a(t'')dt'' \int_{\Delta_{t',t}} d\boldsymbol{r}_1 \int_{\Delta_{t'',t}} d\boldsymbol{r}_2 \, g_2(\boldsymbol{r}_1, \boldsymbol{r}_2, t', t'')\right]. \quad (A3)$$

Eqn.A3 is the probability that no dots of the class $t'$, whose number density is $I_a(t')dt'$, is located within the domain $\Delta_{t',t}$ (with $0 < t' < t$). In eqn.A3 each class is uniquely identified by $t'$. In modeling phase transformations, dots are nuclei, $t'$ is the birth time of the nucleus and $t - t'$ the age of the nucleus. To determine the volume fraction of the new phase, at time $t$, $\Delta_{t',t}$ is set equal to the volume of the nuclei that start growing at time $t'$.

*B- Expression of $F[Q_a]$.*

In terms of dimensionless units, the integral expression in eqn.9 is given by,

$$F[Q_a] = -(D+1)X_{ex}\left(\int_0^1 Q_a(\tau_1)\,y^D(\tau_1)\,d\tau_1\right)\left(1 + \frac{1}{2}(D+1)X_{ex}\int_0^1 Q_a(\tau_1)\,y^D(\tau_1)\,d\tau_1\right)$$

$$+ \frac{(D(D+1))^2}{\Omega_D}X_{ex}^2\int_0^1 Q_a(\tau_1)d\tau_1\int_0^{\tau_1}Q_a(\tau_2)d\tau_2\int_0^{\tilde{\rho}(\tau_1)}x^{D-1}[H_1(\varphi_0(x)$$

$$+ \varphi_1(x,\tau_2)) + H_2\varphi_2(x,\tau_1,\tau_2)]dx, \qquad (B1)$$

where

$$y^D(\tau_1) = H(1-\rho_{hc})[H(\tau_1+\rho_{hc}-1)\rho_{hc}^D + H(1-\tau_1-\rho_{hc})(1-\tau_1)^D]$$
$$+ H(\rho_{hc}-1)\rho_{hc}^D, \qquad (B2)$$

$$\varphi_0(x) = \omega_D[\rho_{hc},\rho_{hc};x], \qquad (B3)$$

$$\varphi_1(x,\tau_2) = H\big(x-(\rho_{hc}-1+\tau_2)\big)\Big[H(1-\tau_2-\rho_{hc}-x)A(1-\tau_2,\rho_{hc})$$
$$+ H\big(x-(1-\tau_2-\rho_{hc})\big)\omega_D[1-\tau_2,\rho_{hc};x]\Big], \qquad (B4)$$

and

$$\varphi_2(x,\tau_1,\tau_2) = \big[H(1-x-\tau_1)A(1-\tau_2,\tau_1-\tau_2)$$
$$+ H(x+\tau_1-1)\omega_D[1-\tau_2,\tau_1-\tau_2;x]\big]. \qquad (B5)$$

In the equations, $\rho_{hc} = \frac{R_{hc}}{at} = \frac{1}{\bar{t}}\left(\frac{\gamma}{3}\right)^{1/(D+1)}$ is the reduced radius of the correlation sphere where $\bar{t}$ is the reduced actual time and $\gamma = R_{hc}^{D+1}\frac{\pi I_0}{a}$. Eqns.B3,B4 refer to the case $\tau_1 - \tau_2 < \rho_{hc}$ (i.e. $R(t_1 - t_2) < R_{hc}$) and eqn.B5 to the case $\tau_1 - \tau_2 > \rho_{hc}$ (i.e. $R(t_1 - t_2) > R_{hc}$). Accordingly, in eqn.B1, $H_1 \equiv H(\rho_{hc} - (\tau_1 - \tau_2))$ and $H_2 \equiv 1 - H_1$. As regards the extreme of integration for $x$, namely $\tilde{\rho}(\tau_1)$, and the integration domains for $\tau_1$, and $\tau_2$ in eqn.B1, we must refer to the three cases outlined in the text (see also Fig.3). To this end, the integration domains of the time variables are considered in eqn.B1 by multiplying the $Q_a$ integrand for the Heaviside step functions reported below, together with the $\tilde{\rho}(\tau_1)$ extreme of integration:

Case i)   $1 - \rho_{hc} < \tau_2 < \tau_1$ i.e.  $H(\tau_2 - (1 - \rho_{hc}))$ and $\tilde{\rho}(\tau_1) = \rho_{hc}$

Case ii)   $\tau_2 < 1 - \rho_{hc} < \tau_1$ i.e. $H((1-\rho_{hc}) - \tau_2) H(\tau_1 - (1-\rho_{hc}))$ and $\tilde{\rho}(\tau_1) = \rho_{hc}$

Case iii)  $\tau_2 < \tau_1 < 1 - \rho_{hc}$ i.e. $H((1-\rho_{hc}) - \tau_1)$ and $\tilde{\rho}(\tau_1) = 1 - \tau_1$

As stressed in the main text, for $\tau_1 - \tau_2 > \rho_{hc}$ (i.e. $R(t_1 - t_2) > R_{hc}$) only cases ii) and iii) hold.

In eqns.B4,B5, $A(\rho_a, \rho_b) = \frac{\Omega_D}{D}[\rho_a^D - \rho_b^D]$ and $\omega_D[\rho_a, \rho_b; x]$ is the measure of $\Delta(\rho_b)\backslash\Delta(\rho_a)$, i.e. the volume of the sphere $\Delta(\rho_b)$ minus the overlap volume between $\Delta(\rho_a)$ and $\Delta(\rho_b)$. For $D = 2$ and $D = 3$ these volumes are given by:

$$\omega_2[\rho_b, \rho_a; x] = \rho_b^2 \left[\pi - \cos^{-1}\frac{x^2 - (\rho_a^2 - \rho_b^2)}{2x\rho_b}\right] - \rho_a^2 \cos^{-1}\frac{x^2 + (\rho_a^2 - \rho_b^2)}{2x\rho_a}$$
$$+ \frac{1}{2}\sqrt{4x^2\rho_b^2 - [x^2 - (\rho_a^2 - \rho_b^2)]^2}$$

and

$$\omega_3[\rho_b, \rho_a; x] = \pi\left[\frac{2}{3}(\rho_b^3 - \rho_a^3) - \frac{1}{12}x^3 + \frac{1}{2}x(\rho_a^2 + \rho_b^2) + \frac{(\rho_a^2 - \rho_b^2)^2}{4x}\right].$$

In these equations $\rho_k = \frac{R_k}{at}$ is the reduced radius.

Eqn.9 in the main text was solved numerically by successive iteration, $Q_a^{(k)} = \exp\left(F[Q_a^{(k-1)}]\right)$, starting from $Q_a^{(0)}(\bar{t}) = e^{-X_{ex}(\bar{t})} = e^{-\bar{t}^{(D+1)}}$ that is the KJMA solution. Fig.B1 shows the behavior of the relative integral error of the iterative process, $\eta_k$, as a function of the number of iterations, $k$:

$\eta_k = 2\frac{\int [Q_a^{(k)} - Q_a^{(k-1)}]d\bar{t}}{\int [Q_a^{(k)} + Q_a^{(k-1)}]d\bar{t}}$. At $k = 9$ the error is lower than $10^{-4}$ and the two last kinetics (at $k = 9$ and $k = 8$) are practically indistinguishable in the entire time domain (not shown).

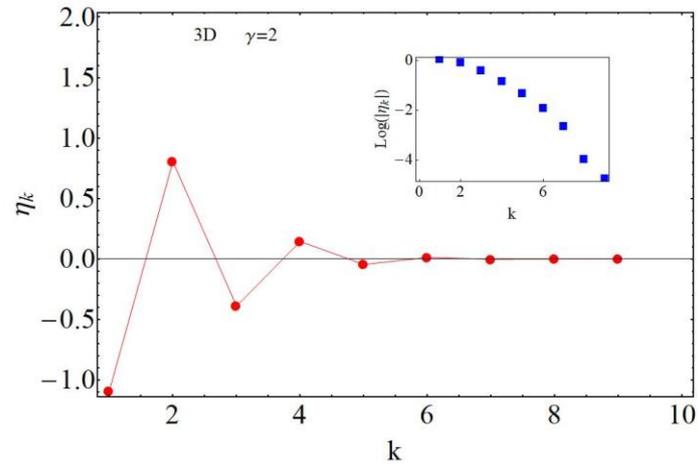

Fig.B1 Behavior of $\eta_k$ vs $k$ up to $k = 9$, for $D = 3$ and $\gamma = 2$ (solid circles). The solid line is a guide for the eyes. Inset: the decimal log of the absolute value of $\eta_k$ is plotted as a function of $k$.